\documentclass{aa}
\usepackage{graphicx}
\usepackage{epsfig}
\usepackage{txfonts}
\usepackage{longtable,lscape}
\usepackage{aalongtable}

\begin{document}

\title{A revised catalogue of EGRET $\gamma$-ray sources}

\author{Jean-Marc Casandjian \inst{1} \and Isabelle A. Grenier \inst{1}   }

\offprints{J.-M. Casandjian}
    
\institute{Laboratoire AIM, CEA/DSM - CNRS - Universit\'e Paris Diderot,\\
Service d'Astrophysique, CEA Saclay, 91191 Gif sur Yvette, France \\
\email{casandjian@cea.fr \& isabelle.grenier@cea.fr} }

\date{Received ; accepted }

\abstract
{}
{We present a catalog of point $\gamma$-ray sources detected by the EGRET detector aboard 
the Compton Gamma Ray Observatory. We have used the whole $\gamma$-ray dataset of reprocessed photons 
at energies above 100 MeV together with new Galactic interstellar emission models based on 
recent $CO$, $HI$, dark gas, and interstellar radiation field data. Two different assumptions have been used for the cosmic-ray distribution in the Galaxy to explore the resulting systematic uncertainties in source detection and characterization.}
{We have used the same 2-dimensional maximum-likelihood detection method as for the 3rd EGRET catalogue.}
{The revised catalogue lists 188 sources, 14 of which are marked as confused, compared to the 271 entries of the 3rd EGRET (3EG) catalogue. 107 former sources have not been confirmed because of the additional structure in the interstellar background. The vast majority of them were unidentified and marked as possibly extended or confused in the 3EG catalogue. In particular, we do not confirm most of the 3EG sources associated with the local clouds of the Gould Belt. Alternatively, we find 30 new sources with no 3EG counterpart. The new error circles for the confirmed 3EG sources largely overlap the previous ones, but several counterparts of particular interest that had been discussed in the litterature, such as Sgr A*, radiogalaxies and several microquasars are now found outside the error circles. We have cross-correlated the source positions with a large number of radio pulsars, pulsar wind nebulae, supernova remnants, OB associations, blazars and flat radiosources and we find a surprising large number of sources (87) at all latitudes with no counterpart among the potential $\gamma$-ray emitters.}
{}

\maketitle

\section{Introduction}
The Energetic Gamma-Ray Experiment Telescope (EGRET), which
operated on board the Compton-Gamma Ray Observatory from April
1991 to May 2000, detected photons in the 20 MeV to 30 GeV range.
The observation program made use of the large instrumental field
of view (25$^{\circ}$ in radius) to cover the whole sky and for in-depth
studies of specific regions. The resulting exposure and flux
sensitivity to point sources are therefore not uniform across the
sky. The sensitivity threshold also varies because of the intense
background emission that arises from cosmic-ray interactions with
the interstellar gas and photon fields in the Milky Way. The
minimum flux that EGRET could detect steeply rises with decreasing
Galactic latitude. In order to detect point sources and assess
their significance in these varying conditions, a 2-dimensional 
maximum-likelihood method using binned maps had been developed for
the COS-B data (\cite{pollock81}) and implemented for the EGRET
one (\cite{mattox96}). A first catalog using this method was
published after 1.5 years of data (\cite{fichtel94}), followed by
the second one (\cite{thompson95}) and its supplement
(\cite{thompson96}) after 3 years of data. Lamb \& Macomb (1997)
presented a catalog of sources detected above 1 GeV. The last
EGRET catalog (hereafter 3EG, \cite{hartman99}) comprised
reprocessed data from April 1991 to October 1995 with the
interstellar emission model from Hunter et al. (1997) and
extragalactic background from Sreekumar et al. (1998). This version
contained 271 point sources including a solar flare, the Large
Magellanic Cloud, five pulsars, one radiogalaxy detection (Cen
A), 66 high-confidence identifications of blazars (BL Lac objects and
flat-spectrum radio quasars), and 27 lower-confidence blazar
identifications. Because of the wide tails of the instrument
point-spread function, seven potential artifacts were noted around
the brightest sources and many sources were marked as confused or
possibly extended.

The 3EG catalogue also contained 170 sources with no attractive
counterpart at lower energy. About 130 of them remain unidentified
as of today (see Grenier (2004) and references therein). Candidate
counterparts that have been searched for include pulsars and their
wind nebulae, supernova remnants, massive stars, X-ray binaries
and microquasars, blazars and nearby radiogalaxies, luminous
infrared and starburst galaxies, and galaxy clusters. It was also
noticed (\cite{grenier95,grenier00,gehrels00}) that the most
stable unidentified sources are significantly correlated with the
nearby Gould Belt, a system of massive stars and interstellar
clouds that surrounds the Sun at a distance of 
hundreds of parsecs. The offset position of the Sun with respect to the Belt centre 
and the Belt inclination of 17$^{\circ}$ to the Galactic plane indeed 
provides a useful spatial signature across the sky (\cite{perrot03}).

EGRET went on observing for another 4.5 years after the 4 cycles
used for the 3EG work. Its sensitivity was reduced because of the
ageing gas in the spark chamber, but it gathered nearly ten
percent more photons and saw several new variable sources. Several
authors (\cite{nolan03,sowards05}), however, noticed discrepancies
between their studies and at least five 3EG sources. They failed to
confirm sources and found others. The whole $\gamma$-ray dataset and
final instrument response functions have also been significantly
reprocessed by the EGRET team in 2001. Furthermore, the spatial
coverage of the $CO$ surveys has reached higher latitudes since
1999, finding new small $CO$ clouds (\cite{dame01}). In parallel,
new $HI$ surveys (\cite{kalberla05}) have been completed to correct
for the significant contamination of stray radiation in the older
ones. Finally, an additional 'dark' gas component has been found
in the Gould Belt clouds that significantly increases their mass
and spatial extent (\cite{grenier05}). The additional mass is
structured into large envelopes around the dense $CO$ cores. They do
not follow the $HI$ and $CO$ maps commonly used to trace atomic and
molecular column-densities. So, the dark gas provides both
$\gamma$-ray intensity and structure that were not accounted for in
the 3EG background model.

For all these reasons and in preparation of the new GLAST mission,
it was necessary to revise the interstellar background model and
to apply the EGRET detection method to the full nine years of data
to build a new catalogue of sources above 100 MeV. In order to
study the systematic uncertainties induced on source locations and
fluxes by our limited knowledge of the intense interstellar
background, we have applied the analysis to two different
background models exploiting the same new interstellar data, but
using different approaches to constrain the cosmic-ray gradient across the Galaxy.

\section{The Galactic interstellar emission models}

\begin{figure*}
\resizebox{\hsize}{!}
{\includegraphics{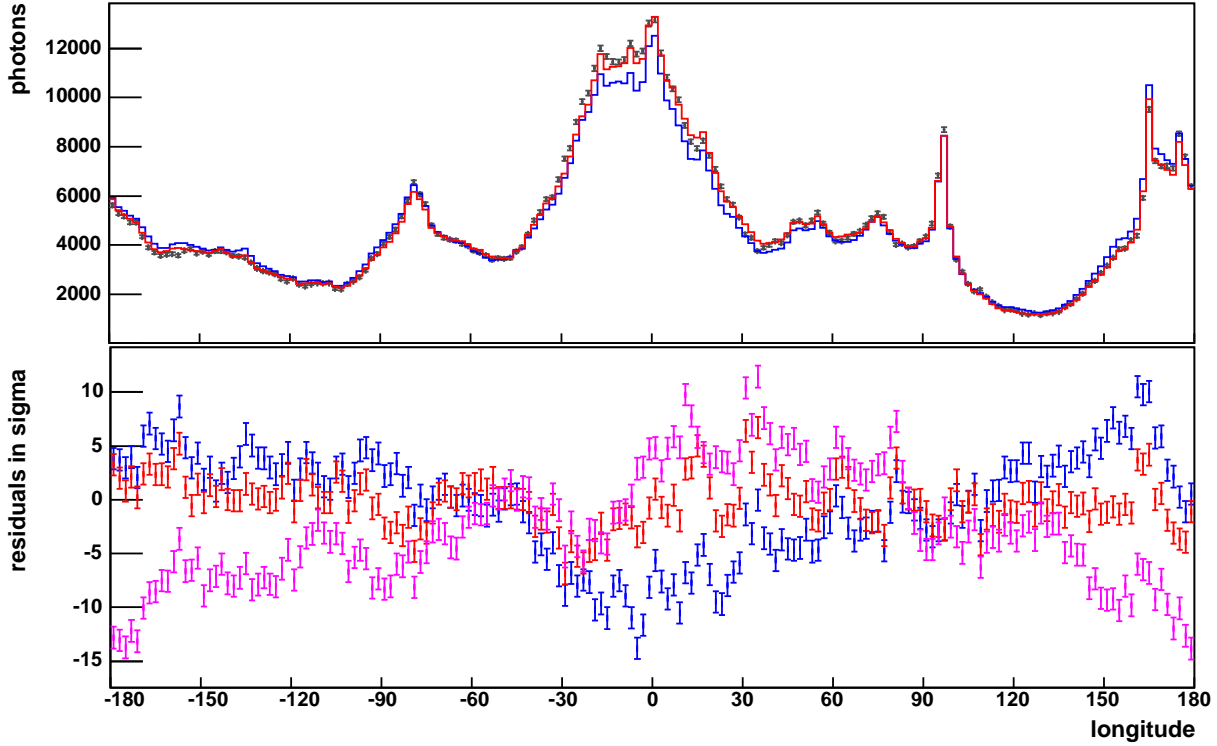}}
\centering
\caption{The top figure is the longitude profile of all photon counts observed 
by EGRET above 100 MeV at all latitudes (black error bars), compared with the diffuse counts
predicted by the 3EG model (blue curve) and the Ring model (red curve). 
The bottom figure is the residual expressed in number of standard deviation, colors are the 
same as above, we added the Galprop residuals in purple. Counts from bright sources have been added to 
the diffuse component. For more visibility the plot are presented with a 
binning of 4$^{\circ}$.}
\label{Figlongitude}
\end{figure*}

The high-energy Galactic emission is produced by the interaction
of energetic cosmic-ray electrons and protons with interstellar
nucleons and photons. The decay of neutral pions produced in
hadron collisions accounts for most of the emission above 300 MeV.
Inverse Compton (IC) scattering of the interstellar radiation
field by electrons and their Bremsstrahlung emission in the
interstellar gas are the other main contributors to the Galactic
emission. The observed intensity therefore scales with the
integral along the line of sight of the cosmic-ray density times
the gas or soft-photon one.

The diffuse model used for the 3EG catalogue (\cite{hunter97}) was
based on a 3D-distribution of matter, cosmic-ray and soft-photon
densities in the Galaxy, where the cosmic-ray density was assumed
to be coupled to the gas one over a given length scale. This
length as well as the $CO$-to-$H2$ conversion factor (X ratio) were
adjusted to the data. The 3D gas map was obtained from the $HI$ and
$CO$ line surveys and from kinematical distances derived for circular
rotation. Distance ambiguities in the inner Galaxy were solved by
splitting the gas into the far and near sides according to its
expected scale height. Gas with velocities in excess of the
tangent values was attributed to the tangent point and gas
emission within 10$^{\circ}$ of the Galactic center and anticenter was
interpolated from the regions just outside these boundaries and
normalized to match the total emission seen along the line of
sight. The resulting map is, however, still strongly biased to our
side of the Galaxy, particularly for the atomic gas. This bias is
reflected in the cosmic-ray density via the coupling length.

For the present analyses, we have assumed an axisymmetric Galaxy
for the cosmic-ray density and we have used gas column-density
distributions in Galactocentric rings that are less subject to
biases due to the strategy adopted to solve the cloud distance in
the inner Galaxy. The radial velocity information in the $HI$ and $CO$
line surveys, together with the rotation curve of Clemens (1985)
and the solar motion (v  = 220 km/s at R  = 8.5 kpc), have been
used to partition the gas into 6 rings bounded by 3.5, 7.5, 9.5,
11.5, and 13.5 kpc in Galactocentric distance (Digel et al., in
preparation). Gas within 10$^{\circ}$ of the Galactic center and anticenter
was interpolated as before. The all-sky Leiden-Argentina-Bonn
(LAB) composite survey (\cite{kalberla05}) was used for the $HI$
data. Column densities, $N(HI)$, were derived under the assumption
of a constant spin temperature of 125 K. The velocity-integrated
$CO$ brightness temperature, $W(CO)$, comes from the Center for
Astrophysics compilation of observations at $|b| \leq 32 ^{\circ}$
(\cite{dame01}). The regions outside the survey boundaries should
be free of bright $CO$ emission.

We have used two different approaches to account for the
cosmic-ray density gradient. One is based on the Galprop model for
cosmic-ray propagation developed by Strong et al. (2007, 2004a, 2004b), 
using run number 49-6002029RB to derive the
$\gamma$-ray maps from pion decay, $I_{\pi^0}$, bremstrahlung
radiation, $I_{brem}$, and inverse Compton radiation, $I_{IC}$.
This version includes secondary electrons and positrons, an
optimized cosmic-ray spectrum to fit the GeV excess in the EGRET
data, a cosmic-ray source distribution matching the radial profile
of pulsars and supernova remnants, a radial gradient in the X
factor, and the new $HI$ and $CO$ gas rings.

The second model, hereby referred to as the Ring model, is based
on the simpler, but realistic hypothesis that, if energetic cosmic
rays uniformly penetrate all gas phases, the $\gamma$-ray intensity
in each direction can be modelled as a linear combination of gas
column-densities in the different rings, plus the IC intensity map
(as predicted by Galprop), and an isotropic intensity ($I_{iso}$)
that accounts for very local IC emission and extragalactic
emission. This assumption has been used to derive gas emissivities
in several rings from the COS-B and EGRET data
(\cite{strong88,strongmattox96}). We have reproduced these
analyzes to derive gas emissivities for the new $HI$ and $CO$ rings
using 9 years of EGRET data in three energy bands ($> 100$ MeV,
$0.3-1$ GeV, $> 1$ GeV). Both the Ring and Galprop models used the
revised distribution of the interstellar radiation field
(\cite{porter05,moskalenko06}) to calculate the IC intensity map.
The Galprop IC map is common to both diffuse models.

As indicated in the introduction, we have also included in the
local ring the large column-densities of "dark" gas associated
with cold and anomalous dust at the transition between the atomic
and molecular phases (\cite{grenier05}). This transitional phase
is not traced in the radio. When removing from total dust
column-density maps the part that linearly correlates with $N(HI)$
and $W(CO)$, one is left with large envelopes of excess dust
around all the nearby $CO$ clouds. The fact that the excess dust
spatially correlates with significant diffuse gamma radiation
indicates that cosmic rays pervade gas not accounted for in $HI$ or
$CO$. The gas-to-dust ratio in this phase, as inferred from the
excess dust and correlated $\gamma$-ray data, is normal. This phase
appears to form an extended layer at the transition between the
dense $CO$ cores and the densest parts of the outer $HI$ envelope of a
cloud complex. It is best seen in total dust maps such as the
reddening E(B-V) map (\cite{schlegel98}), or low-frequency thermal
emission at 93 GHz for WMAP (\cite{finkbeiner99}), or anomalous
emission near 20 GHz (\cite{lagache03}). We constructed a "dark"
gas column-density template, $NH_{dark}$, by removing from the
E(B-V) map the part linearly correlated with $N(HI)$ and $W(CO)$. This
template was turned into gas column-densities by fitting it
together with the $N(HI)$ and $W(CO)$ rings, as well as IC and
isotropic components, to the all-sky $\gamma$-ray maps. Because of
its column-densities, clumpiness, and large spread across the sky
(see Figure 4 in Grenier et al. (2005)), the "dark" gas component
may strongly affect source detectability. This template was also
added to the Galprop 49-6002029RB background model.

To summarize, two diffuse backgrounds were constructed by fitting
different components to the EGRET photon maps, in $0.5^{\circ}
\times 0.5^{\circ}$ bins, in the three energy bands that will be
used for source detection ($> 100$ MeV, $0.3-1$ GeV, $> 1$ GeV).
\begin{enumerate}
\item With the Ring model, the predicted count rates are
calculated as:
\begin{equation}
\begin{array}{l}
 N_{pred}(l,b) = [\sum_{i=rings} q_{HI,i} N_{HI}(r_i,l,b) + \sum_{rings} q_{CO,i} W_{CO}(r_i,l,b) \\

 + q_{dark} NH_{dark}(l,b) + q_{IC} I_{IC}(l,b) + I_{iso}] \times \epsilon(l,b) \\

 + \sum_{j=sources} \epsilon(l_j,b_j)\, f_{j}\, PSF(l_j,b_j)
\label{eqRing}
\end{array}
\end{equation}

\item and the Galprop model as:
\begin{equation}
\begin{array}{l}
 N_{pred}(l,b) = [q_{\pi^0} I_{\pi^0}(l,b) + q_{brem}
 I_{brem}(l,b) + q_{dark} NH_{dark}(l,b) \\

 + q_{IC} I_{IC}(l,b) + I_{iso}] \times \epsilon(l,b) \\

 + \sum_{j=sources} \epsilon(l_j,b_j)\, f_{j}\, PSF(l_j,b_j)
\label{eqGalprop}
\end{array}
\end{equation}
\end{enumerate}

In both models, $\epsilon(l,b)$ and $f_{j}$ note the EGRET exposure map and source fluxes. The
diffuse maps times the exposure were convolved with the EGRET PSF
for an input $E^{-2.1}$ spectrum before adding the source maps.
The EGRET count and exposure maps, the 3EG diffuse model, as well
as the latest instrument response functions, were downloaded from
the CGRO Science Support Center. They differ from those used for
3EG since they were reprocessed in 2001. The $q$ parameters (gas
emissivities or relative contributions of different radiation
components) were fitted to the data by means of a maximum
likelihood with Poisson statistics. To avoid biasing the
interstellar parameters, the model included the brightest sources 
detected during a first source detection
iteration with a significance $> 5 \sigma$, with fixed fluxes.
Changing these fluxes within their statistical uncertainties do
not significantly change the diffuse results.

The resulting emissivities corresponding to the local gas are 
fully consistant with Grenier et al. (2005) Table 1. 
The emissivity gradient in the Galactic plane will be described 
in a separate paper. The quality of the fit can be seen in Figure \ref{Figlongitude}.
The top figure displays the longitude profile of all the EGRET photon
counts above 100 MeV. The error bars are only statistical. The
plot compares the best fit that can be obtained using the former
3EG diffuse model with the longitude profile resulting from the
present Ring model. The bottom plot shows the longitude profile of the 
residuals and the improvement of the ring model over the 3EG one. 
It also shows the residuals for the best fit Galprop model. All
modelled profiles include the brightest sources. Systematic
differences can be seen in various places where the 3EG model
significantly over-predicts and under-predicts the data while the
new models behave better. Because of its larger flexibility (the
gas emissivity gradient due to cosmic-ray variations is measured,
not inferred from propagation properties or gas coupling), the
Ring model was found to best fit the data. It is worth noting than
even if the agreement is excellent, there still exists small
deviations that can significantly impact source detection and
characterization.

\begin{figure}
\resizebox{\hsize}{!}{\includegraphics{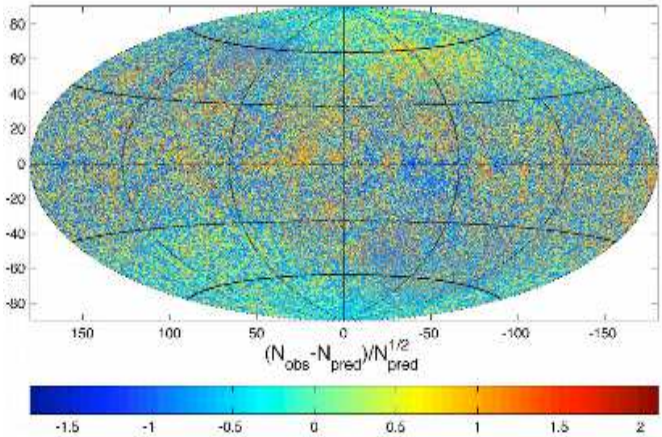}} \centering
\caption{Map in Galactic coordinates of the residuals (expressed
in $\sigma= \sqrt{N_{pred}}$ values) between the $E > 100$ MeV photon counts (in 0.5$^{\circ}$ bin) and
the best fit with the Ring model using Equation (\ref{eqRing}) }.
\label{Figresidual}
\end{figure}

\begin{table*}
\caption{List of individual or short periods used in the analysis
in addition to the summed cycles.} \label{Tabperiods} \centering
\begin{tabular}{l l l l l l}     
\hline\hline Name & Sum of viewing periods & Name & Sum of
viewing periods & Name & Sum of viewing periods\\
2+ & 0002+0003+0004+0005 & 2040 & & 3315\\
0020 & & virg2 & 2040+2050+2060 & 330+ & 3300+3320\\
0040 &  & 2110 &  & 335+ & 3350+3355\\
0050 &  & 2230 &  & vrg3a & 3040+3050+3060+3070+3080+3086\\
0200 &  & 2260 &  & 3355\\
0210 &  & 227+ & 2270+2280 & 3360\\
0220 &  & 229+ & 2290+2295 & 3385\\
0230 &  & 2310 &  & 3390\\
0250 &  & 3023 &  & 4040\\
0260 &  & 314+ & 3140+3150 & 4100\\
0290 &  & 3170 &  & 4130\\
36+ & 0360+0365 & 319+ & 3190+3195 & 4180\\
0420 &  & 3200 &  & 419+ & 4191+4195\\
0430 &  & 328+ & 3280+3310+3315+3330 & 4210\\
0440 &  & 3290 &  & 4230\\
     &  &  &  & 4235\\
\hline
\end{tabular}
\end{table*}

The residual count map obtained above 100 MeV with the Ring model
is presented in Figure 2. It displays the statistical difference
$(N_{obs}-N_{pred})/ \sqrt{N_{pred}}$ between the observed counts and
those predicted from the diffuse background and bright sources using
equation \ref{eqRing}. The model globally fits very well the data.
The extended blue fan-like structures with negative residuals are
correlated with the edge of several observing periods. They
probably result from a wrong exposure estimate at large angle from
the instrument axis. They are visible independently of the choice
of diffuse model (Ring, Galprop, or 3EG). Their spatial extent
is large enough compared to the PSF size not to severely affect
source detection, yet source fluxes in these directions are
underestimated. Uncertain knowledge of the off-axis instrument
exposure is also reflected in the small model deficit (orange
edge) bordering the fan-like excesses. We have checked for
suspicious strings of faint sources that would correlate with
these instrumental features. 

The use of two different background models allowed us to study
their impact on source detection and characterization. Given its
higher likelihood value and locally flatter residuals, the Ring model
was used to derive the default source flux and location. The
values obtained with the Galprop background are used to
illustrate the amplitude of the systematic uncertainty due to the
background modelling. When searching for sources we used the diffuse emission 
parameters calculated from this global fit. We adjusted a source flux together 
with a free normalization of the total diffuse flux within 15$^{\circ}$ around each 
pixel, and a free isotropic flux. This procedure is the same as used for 
3EG (Gmult and Gbias). These two parameters correct for small local 
mismatches between the diffuse model and the data. Gmult fluctuates around 1.

\section{Source detection}

As for the derivation of the 3EG catalogue, we have used the LIKE
code (Mattox, 1996, version 5.61) to compute the 2-dimension binned Poisson
likelihood of detecting a source at a particular location on top
of the diffuse background. LIKE calculates the Test Statistic ($TS$)
value that compares the likelihood of detecting a PSF-like excess
above the background to the null hypothesis - a random background
fluctuation - for a given position. The likelihood ($L_{i}$) is calculated
as the product, for all pixels within 15$^{\circ}$ of a specific position,
of the Poisson probabilities of observing photons in a pixel
where the number of counts is predicted by the model (background + source). 
The likelihood ratio test statistic is defined as $TS = -2 (Ln L_{0} - Ln
L_{1})$, where the likelihood values $L_{1}$ and $L_{0}$ are respectively
optimized with and without a source in the model. Asymptotically,
the $TS$ distribution follows a  $\chi^{2}$ one. The detection significance of a 
source at the given position is $\sqrt{TS} \sigma$  (Mattox 1996).

Sources have been searched for in the summed maps corresponding to
cycle 1, 2, 3, 4, 1+2, 3+4, 1+2+3+4, 5+6, 7+8+9,
1+2+3+4+5+6+7+8+9. In addition, we have analyzed the 46 individual
periods listed in Table \ref{Tabperiods} for which flaring 3EG sources had been
detected. As for the summed maps, the individual period maps
retained only photons with inclinations within 30$^{\circ}$ from the
instrument axis, or $19^{\circ}$ for cycle 6, 7, 8, and 9. Photons
and exposure maps were binned to $0.5^{\circ} \times 0.5^{\circ}$.

To build the 3EG catalogue, sources were detected only in the
integrated $E > 100$ MeV band. $TS$ maps were then constructed in
three energy bands ($> 100$ MeV, $0.3-1$ GeV, and $> 1$ GeV) from
the observation (single or summed) with highest  $TS$ and a source
final position was obtained from the smallest error contours.
Given the modern computer performance, we have directly searched
for sources independently in the three energy bands.

At 100 MeV, the EGRET PSF is wide and there exists discrepancies
between its real shape, as observed in bright sources, and the
modelled one. In practice, differences may also come from a more
complex source spectrum than the single power-law assumed to
integrate the PSF. A choice of 300 MeV instead of 100 MeV for the
lower analysis threshold might have been a better trade-of between count
rates for detection and systematic uncertainties in the PSF. 
We have, however, kept a lower limit of 100 MeV as in 3EG in order to 
account for soft sources and to allow comparison with the 3EG results. 
We have assumed a spectral index of 2.0 for all sources but for 11
bright ones which had a 3EG spectral index far from 2.0. For the
latter, we have used their 3EG index to integrate the PSF.

Each of the 10 all-sky summed maps was divided, both in Galactic
and equatorial coordinates, in 45 zones with a large overlap. The
use of both coordinates systems is required since source images
are deformed in rectangular projection at high latitude or
declination. For each zone, each individual period, and each of
the 3 energy bands ($>100$ MeV, $0.3-1$ GeV, and $>1$ GeV), we
calculated a $TS$ map for excesses above the background. Sources
were iteratively detected from high $TS$ to low $TS$ in successive $TS$
maps. Between each steps, the detected sources were included in
the background model until no excess with  $\sqrt{TS} > 3$ was left in
the final $TS$ map. An example of the iteration around Geminga is
given in Figure 3. Peaks in the $TS$ map were automatically detected
with SExtractor (\cite{bertin96}) and converted into source
position by taking the $TS$-weighted centroid in the region enclosed
by the 95\% confidence contour around this position. Source
positions were recalculated at each iteration to take into account
the influence of the neighbouring sources. More than 1100 $TS$-maps
were thus calculated at the CCIN2P3 Computing Center.

\begin{figure}
\centering
\resizebox{\hsize}{!}{\includegraphics{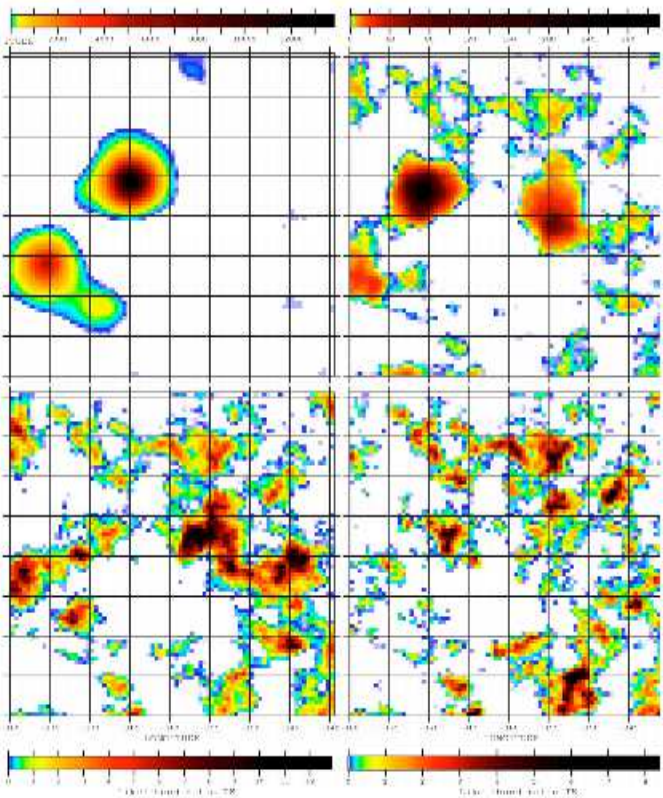}} \centering
\caption{An example of the iterative source detection with the 2D
binned likelihood around Geminga at energies above 100 MeV. 4 consecutive $TS$ maps are shown.
Sources are detected, then are included in the background for the
next step until no significant one is left. The colourbar gives
$TS$.} \label{FigTSiteration}
\end{figure}

\section{Catalogue construction}
To account for real versus modelled PSF discrepancies in extremely
bright sources, for instance to account for the splitting in two
of the bright pulsar sources or for the artifacts in the Vela
tails, we have removed all the source candidates within
$3.5^{\circ}$ of the intense sources (that exhibit more than 800
photons in a map).  For less intense sources, we have checked the
probability of having a double versus single source with a
specific likelihood calculation, using the likelihood ratio
between the 2 cases to keep or reject the double source.

At the end of this stage, most sources have two possible positions per energy
band and observation, one from the Galactic coordinate map and one
from the equatorial one. We cross-compared the two and selected
the position from the least deformed projection. Sources detected
only once were not included in the list unless their latitude
or declination were higher than $40^{\circ}$ or their longitude or
right-ascension were less than $5^{\circ}$ from the map edges.

At this stage, most sources have three possible positions (with energy) for a
given observation. We chose among the three the position
corresponding to the smallest 95\% confidence contour, unless its
peak $\sqrt{TS}$ were 1.5 smaller than found in an another energy
band. The latter condition reduces the risk of incorrect source
assignment during the cross-comparison phase. Sources found at low
energy, but not at high energy were included in the list, as well
as sources found only at high energy.

We have used the same criteria to cross-compare the source
positions for individual periods and summed cycles in order to
obtain a final list of candidate sources with the best position
from the different energy bands and periods/cycles. We followed
the whole procedure with both the Ring and Galprop interstellar
backgrounds. We obtained respectively 1192 and 1225 candidate
sources with the Ring and Galprop models. Source fluxes and $\sqrt{TS}$ 
values above 100 MeV were calculated for these sets of
positions for the different periods and cycles. Unlike in 3EG, we
did not adjust the position of the identified sources (AGN or
pulsars) to that of their radio counterpart.

We adopted the same detection threshold as for the 3EG catalogue
($\sqrt{TS} > 5$ at $|b| < 10^{\circ}$ and $\sqrt{TS} > 4$
elsewhere) and found 188 and 208 significant sources for the Ring
and Galprop models, respectively. We manually checked the $TS$ maps
of all the sources that barely passed the detection threshold with
the Ring model and had $\sqrt{TS} \sim 3$ with the Galprop one.

We emphasize the fact that the order and criteria applied to
cross-correlate positions between the excesses detected in
different energy bands and time periods can strongly affect the
catalog list near the detection threshold. Several strategies were
tested before adopting the present one, but one must remember that
a faint source can pass or drop below the threshold by slightly
changing its position or that of its neighbours. Given the steep
increase in source numbers with decreasing $TS$, we also emphasize
that a small change in the $TS$ threshold, alternatively in the
background over which the source $TS$ is calculated, results in a
large change in the number of catalogue entries. For instance,
lowering the $\sqrt{TS}$ threshold by 0.1 would add 27 sources.

\section{Catalogue description}
The EGR acronym has been adopted for the EGret Revised source list
presented in Table \ref{TabEGR} and Table \ref{TabEGR_full} in a format similar to the 3EG
one. As explained above, the source characteristics (position and
flux, and their uncertainties) have been determined with the Ring
model because of its higher flexibility, better fit, and flatter
residual map. A secondary position and flux has been measured with
the Galprop model and is listed in Table \ref{TabEGR} and Table \ref{TabEGR_full} to
illustrate the amplitude of the systematic uncertainties due to
the choice of interstellar model.

Sources found within a radius of 1.5 PSF FWHM from a very bright
source, and/or with very asymmetric $TS$ map contours are not
included in Table \ref{TabEGR} and Table \ref{TabEGR_full}. Still, they represent significant
excesses of photons above the background which may be due to
extended sources, or structures not properly modelled in the
interstellar emission, or artifacts due to incorrect PSF tails.
This list of 14 confused sources is given in Table \ref{TabEGRc}, 
under the acronym EGRc for EGret Revised confused.

For both tables, the description for each column follows:
\begin{enumerate}
\item Num: source number in order of increasing right ascension.
\item Name: source name based on J2000 coordinates. \item RA and
Dec: J2000 equatorial coordinates in degrees. \item l and b:
Galactic coordinates in degrees. \item $\theta_{95}$: angular
radius, in degrees, of a circular cone which contains the same
solid angle as the 95\% confidence contour. \item F: flux in
$10^{-8}$ photon cm$^{-2}$ s$^{-1}$ for $E > 100$ MeV and for each
time period. \item $\sigma_F$: $1 \sigma$ statistical flux
uncertainty in $10^{-8}$ photon cm$^{-2}$ s$^{-1}$. \item Cnts:
number of photons detected with $E > 100$ MeV. \item $\sqrt{TS}$:
statistical significance of the detection. \item vp: short viewing
period as defined in Table \ref{Tabperiods} or summed cycles noted
$px$ for cycle x, $pijkl$ for the sum of cycles i, j, k, and l,
and $p19$ for the total of 9 cycles. \item l$_{sys}$ and b$_{sys}$ : Galactic
longitude and latitude obtained with the Galprop background model.
\item F$_{sys}$: flux obtained with the Galprop background model, in
$10^{-8}$ photon cm$^{-2}$ s$^{-1}$. \item 3EG: third EGRET
catalog counterpart source name if one exists within a radius of 1
PSF FWHM ($2^{\circ}$ for $E > 100$ MeV) from the EGR source and
if the nearest neighbour relation between the EGR and 3EG sources
is univocal (the nearest neighbour of the EGR source is the 3EG
one and vice versa).
\end{enumerate}

\begin{table*}
\caption{The EGR catalogue. The three first sources are shown. The full catalogue is available with the on-line version} 
\label{TabEGR} 
\centering
\resizebox{17cm}{!}{
\begin{tabular}{l l l l l l l l l l l l l l l l l l l l}     
\hline\hline 
Num & Name       & RA    & Dec   & l      & b     & $\theta_{95}$ & F    & $\sigma_F$ & Cnts & $\sqrt{TS}$ & vp & l$_{sys}$ & b$_{sys}$ & F$_{sys}$ & 3EG      \\
  1 &EGR J0008+7308&   2.01&  73.14&  119.75&  10.54& 0.20&  39.7&   4.4&  330&   10.9&     p19&  119.75&  10.54&   41.0&  3EGJ0010+7309\\
    &             &       &       &        &       &     &  63.9&  11.6&   96&    7.2&      p1&	   &	   &	   &		   \\
    &             &       &       &        &       &     &  33.4&   9.6&   61&    4.1&      p2&	   &	   &	   &		   \\
    &             &       &       &        &       &     &  22.4&   8.7&   37&    3.0&      p4&	   &	   &	   &		   \\
    &             &       &       &        &       &     &  48.8&   7.4&  162&    8.2&     p12&	   &	   &	   &		   \\
    &             &       &       &        &       &     &  21.6&   7.3&   52&    3.4&     p34&	   &	   &	   &		   \\
    &             &       &       &        &       &     &  37.0&   5.3&  212&    8.5&   p1234&	   &	   &	   &		   \\
    &             &       &       &        &       &     &  44.6&   8.3&  115&    6.6&     p56&	   &	   &	   &		   \\
    &             &       &       &        &       &     &  33.1&   9.7&   60&    4.0&    2110&	   &	   &	   &		   \\
  2 &EGR J0028+0457&   7.06&   4.95&  112.15& -57.44& 0.51&  14.3&   4.6&   31&    4.1&     p34&  112.15& -57.44&   14.4&               \\
    &             &       &       &        &       &     &   7.2&   4.8&   13&    1.7&      p1&	   &	   &	   &		   \\
    &             &       &       &        &       &     &  13.9&   5.9&   20&    3.0&      p3&	   &	   &	   &		   \\
    &             &       &       &        &       &     &  14.0&   7.2&   10&    2.7&      p4&	   &	   &	   &		   \\
    &             &       &       &        &       &     &   7.2&   4.8&   13&    1.7&     p12&	   &	   &	   &		   \\
    &             &       &       &        &       &     &  10.7&   3.3&   43&    4.0&   p1234&	   &	   &	   &		   \\
    &             &       &       &        &       &     &  10.4&   3.1&   46&    4.1&     p19&	   &	   &	   &		   \\
    &             &       &       &        &       &     &  24.2&  11.1&   14&    2.9&    3200&	   &	   &	   &		   \\
    &             &       &       &        &       &     &  24.7&  15.6&    6&    2.3&    3360&	   &	   &	   &		   \\
  3 &EGR J0039-0945&   9.75&  -9.75&  112.76& -72.38& 0.27&  13.0&   3.5&   48&    4.8&     p19&  112.65& -72.40&   13.1&  3EGJ0038-0949\\
    &             &       &       &        &       &     &  14.6&   5.8&   23&    3.4&      p1&	   &	   &	   &		   \\
    &             &       &       &        &       &     &  15.7&   5.4&   24&    4.0&      p4&	   &	   &	   &		   \\
    &             &       &       &        &       &     &  14.6&   5.8&   23&    3.4&     p12&	   &	   &	   &		   \\
    &             &       &       &        &       &     &  11.0&   4.5&   22&    3.2&     p34&	   &	   &	   &		   \\
    &             &       &       &        &       &     &  12.3&   3.6&   43&    4.5&   p1234&	   &	   &	   &		   \\
    &             &       &       &        &       &     &  22.2&  17.7&    5&    1.6&    p789&	   &	   &	   &		   \\
\hline
\end{tabular}
}
\end{table*}

\section{Comparison with the 3EG catalogue}

\begin{figure}
\resizebox{\hsize}{!}{\includegraphics{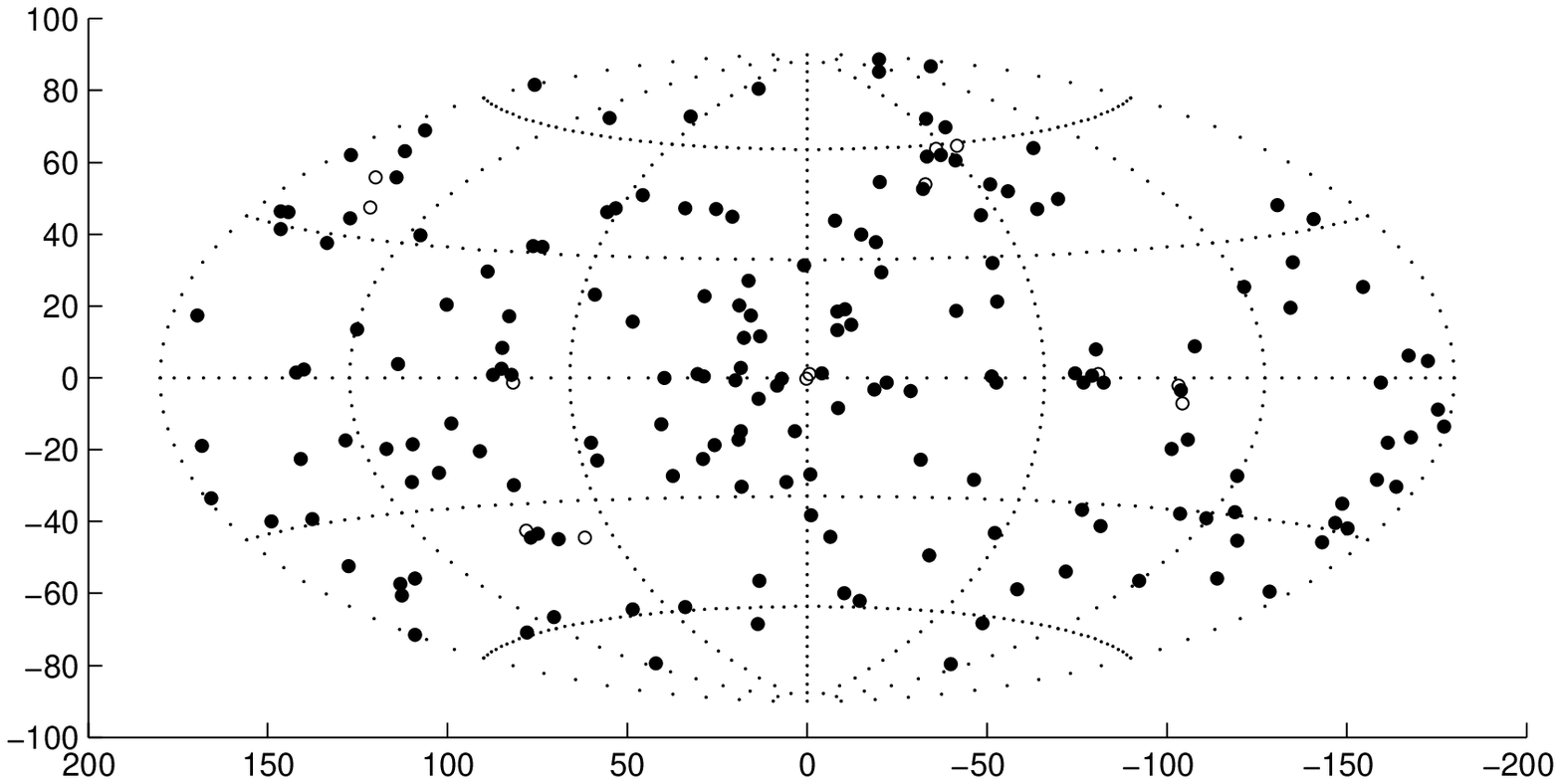}} \centering
\caption{Spatial distribution, in Galactic coordinates, of the EGR
sources. The confused sources are marked as open circles.}
\label{FigskyEGR}
\end{figure}

\begin{figure}
\resizebox{\hsize}{!}{\includegraphics{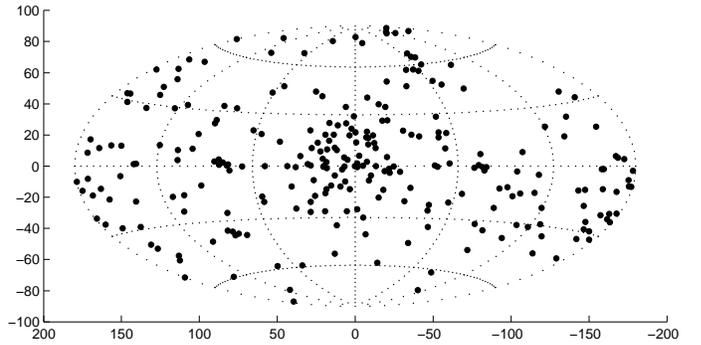}} \centering
\caption{Spatial distribution, in Galactic coordinates, of the 3EG
sources.} \label{Figsky3EG}
\end{figure}

\begin{figure}
\resizebox{\hsize}{!}{\includegraphics{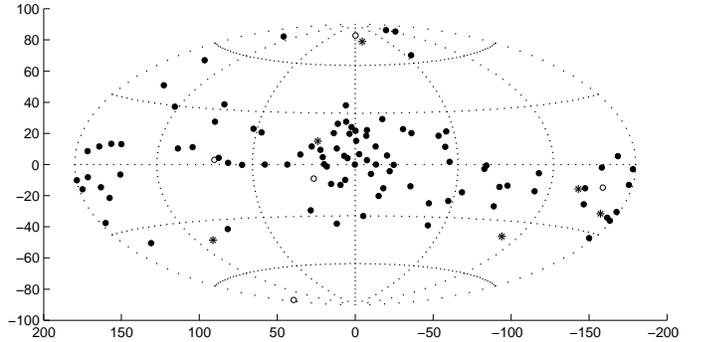}} \centering
\caption{Spatial distribution, in Galactic coordinates, of the 3EG
sources with no counterpart in EGR: the unidentified sources as
circles and the identified AGN as stars. The filled circles and
stars mark the sources that were flagged as extended or confused
in the 3EG catalogue.} \label{Figsky3EGout}
\end{figure}

\begin{figure}
\resizebox{\hsize}{!}{\includegraphics{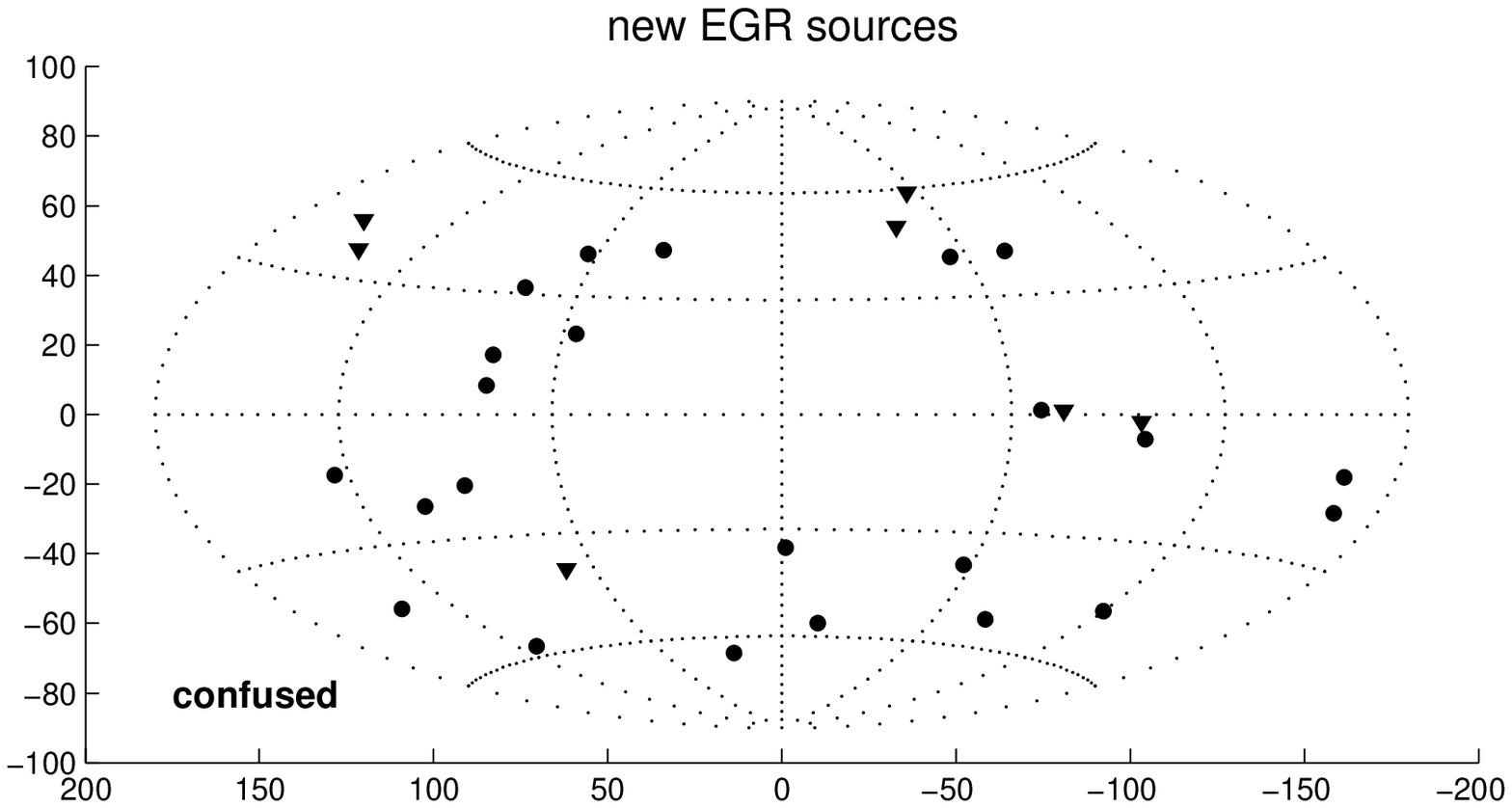}} \centering
\caption{Spatial distribution, in Galactic coordinates, of the new
EGR sources with no 3EG counterpart. The confused sources are
marked as open circles.} \label{FigskyEGRnew}
\end{figure}

\begin{table}
\caption{Names of the 3EG sources with no EGR counterpart}
\label{Tab3EGout} \centering
\begin{tabular}{l l l l l l}     
\hline
3EG J0130-1758 & 3EG J0245+1758 & 3EG J0323+5122\\
3EG J0348+3510 & 3EG J0404+0700 & 3EG J0407+1710\\
3EG J0416+3650 & 3EG J0426+1333 & 3EG J0435+6137\\
3EG J0439+1555 & 3EG J0439+1105 & 3EG J0458-4635\\
3EG J0459+0544 & 3EG J0459+3352 & 3EG J0500+2529\\
3EG J0510+5545 & 3EG J0520+2556 & 3EG J0521+2147\\
3EG J0533+4751 & 3EG J0542+2610 & 3EG J0542-0655\\
3EG J0546+3948 & 3EG J0556+0409 & 3EG J0616-0720\\
3EG J0622-1139 & 3EG J0628+1847 & 3EG J0634+0521\\
3EG J0702-6212 & 3EG J0706-3837 & 3EG J0747-3412\\
3EG J0808-5344 & 3EG J0821-5814 & 3EG J0910+6556\\
3EG J1013-5915 & 3EG J1014-5705 & 3EG J1045-7630\\
3EG J1052+5718 & 3EG J1212+2304 & 3EG J1222+2315\\
3EG J1227+4302 & 3EG J1235+0233 & 3EG J1249-8330\\
3EG J1300-4406 & 3EG J1308+8744 & 3EG J1308-6112\\
3EG J1316-5244 & 3EG J1323+2200 & 3EG J1329+1708\\
3EG J1329-4602 & 3EG J1447-3936 & 3EG J1500-3509\\
3EG J1527-2358 & 3EG J1600-0351 & 3EG J1616-2221\\
3EG J1627-2419 & 3EG J1631-1018 & 3EG J1631-4033\\
3EG J1633-3216 & 3EG J1634-1434 & 3EG J1635-1751\\
3EG J1639-4702 & 3EG J1646-0704 & 3EG J1649-1611\\
3EG J1653-2133 & 3EG J1659-6251 & 3EG J1704-4732\\
3EG J1709-0828 & 3EG J1714-3857 & 3EG J1717-2737\\
3EG J1718-3313 & 3EG J1720-7820 & 3EG J1733+6017\\
3EG J1735-1500 & 3EG J1741-2050 & 3EG J1741-2312\\
3EG J1744-0310 & 3EG J1744-3011 & 3EG J1744-3934\\
3EG J1757-0711 & 3EG J1800-0146 & 3EG J1806-5005\\
3EG J1810-1032 & 3EG J1823-1314 & 3EG J1824+3441\\
3EG J1824-1514 & 3EG J1825+2854 & 3EG J1828+0142\\
3EG J1834-2803 & 3EG J1836-4933 & 3EG J1850+5903\\
3EG J1850-2652 & 3EG J1858-2137 & 3EG J1903+0550\\
3EG J1904-1124 & 3EG J1928+1733 & 3EG J1958+2909\\
3EG J1958-4443 & 3EG J2016+3657 & 3EG J2020-1545\\
3EG J2022+4317 & 3EG J2034-3110 & 3EG J2035+4441\\
3EG J2100+6012 & 3EG J2206+6602 & 3EG J2219-7941\\
3EG J2255+1943 & 3EG J2359+2041\\
\hline
\end{tabular}
\end{table}
\begin{table}
\caption{Names of the new EGR sources with no 3EG counterpart}
\label{TabEGRnew} \centering
\begin{tabular}{l l l}     
\hline
EGR J0028+0457 & EGR J0057-7839 & EGR J0100+4927\\
EGR J0141+1719 & EGR J0243-5930 & EGR J0413-3742\\
EGR J0509+0550 & EGR J0540+0657 & EGR J1122-5946\\
EGR J1158-1950 & EGR J1259-2209 & EGR J1619+2223\\
EGR J1642+3940 & EGR J1740+4946 & EGR J1814+2932\\
EGR J1920+4625 & EGR J1959+4322 & EGR J2027-4206\\ 
EGR J2202+3340 & EGR J2233-4812 & EGR J2258-2745\\
EGR J2308+3645 & EGRc J0818-4613 & EGRc J0842-4501\\ 
EGRc J0912+7146 & EGRc J0927+6054 & EGRc J1038-5724\\
EGRc J1255-0404 & EGRc J1332-1217 & EGRc J2215+0653\\
\hline
\end{tabular}
\end{table}

The revised catalogue contains 174 sources plus 14 confused sources
compared to the 265 entries of the 3EG catalogue (excluding the
Vela artifacts). Their spatial distribution across the sky looks
different from that of the 3EG sources, as illustrated in Figures
\ref{FigskyEGR} and \ref{Figsky3EG}. The accumulation of faint 3EG
sources within $30^{\circ}$ of the Galactic center is much more
reduced in the new results and fewer sources are seen below
$30^{\circ}$ in general. These changes at low and mid latitudes
are primarily due to the increase in background intensity from new
$HI$, $CO$, and dark gas structures. At high latitude, the use of more
$\gamma$-ray observations and of a revised large-scale IC
component in the background may also explain why a handful of 3EG
sources have fallen below the detection threshold whereas new ones
are now detected.

The names of the 107 unconfirmed 3EG sources are listed in Table
\ref{Tab3EGout} and they are displayed in Figure
\ref{Figsky3EGout}. They comprise only six sources that had been
firmly identified as AGN by Hartman et al. (1999), but that had
been flagged as extended or confused by the EGRET team. In fact, the proportion of
these extended or confused cases among the unconfirmed 3EG sources
is overwhelming (95\%) and significantly larger than among the
confirmed ones. The unconfirmed and confirmed 3EG groups
respectively show 69\% and 33\% of possibly extended 'em' sources.
Figure \ref{Figsky3EGout} also shows that the vast majority of
unconfirmed 3EG sources were unidentified and spatially correlated
with the Gould Belt system of nearby clouds. They follow the
characteristic trace of the inclined Belt across the sky,
gathering at $|b| < 30^{\circ}$, more at positive latitudes toward
the Galactic centre and below the plane in the anticenter. The EGR
source sky distribution in Figure \ref{FigskyEGR} does not exhibit
the Gould Belt signature anymore.

\begin{figure}
\resizebox{\hsize}{!}{\includegraphics{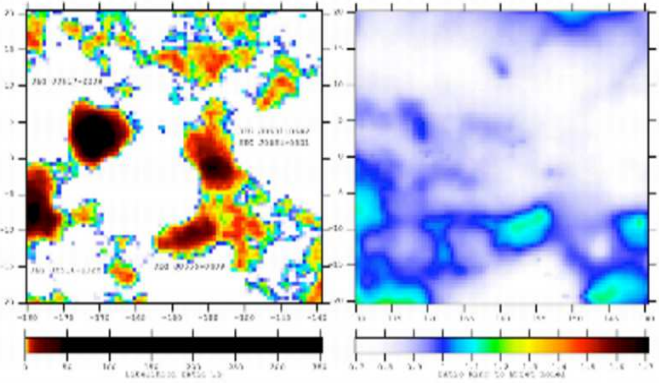}} \centering
\caption{Second stage of the iterative source detection around 
Geminga (see Figure \ref{FigTSiteration}) obtained using the 3EG model (left) 
and map of the Ring model intensity divided by the 
3EG one (right). The excess in the TS map assigned in 3EG to the 3EG J0556+0409 point 
source corresponds to a local underestimation of the diffuse 
emission in the 3EG model. Maps are given in 0.5$^{\circ}$ bins and galactic coordinates}
\label{FigRatioRingEgret}
\end{figure}

The fact that many 3EG sources are not confirmed in the present
analyses should not cast doubts on the detection method from a
statistical point of view. They did correspond to significant
photon excesses above the background in the 3EG analyses, but, in
the absence of some structures in the predicted interstellar
background, an ensemble of point sources with the wide EGRET PSF
would compensate for the missing clouds and yield an excellent fit
to the data. Figure \ref{FigRatioRingEgret} illustrates this point 
with the unidentified source 3EG J0556+0409 detected at 7.2 $\sigma$ in 3EG. 
The left side shows the TS-map corresponding to the second stage of the iterative source detection 
around Geminga above 100 MeV. It is the same as in Figure \ref{FigTSiteration} 
but we have used here the 3EG diffuse emission model instead of the Ring one. The same sources are 
detected except for 3EG J0556+0409 which is not seen in Figure \ref{FigTSiteration}. Instead an 
excess of diffuse emission appears in the ratio of the Ring to 3EG background intensities (Figure \ref{FigRatioRingEgret}, right). 
\emph{The photons attributed to a point source in 3EG where in fact coming from a gas cloud in the Galaxy.}   
This is probably still the case in the present analysis, although to a lesser degree, in particular 
at very low latitude where optical thickness in $HI$ and $CO$ severely limits our
knowledge of the true column-densities. Other sources may also be due to increased cosmic-ray densities in specific clouds 
with respect to the local Galactic average. Over-irradiated clouds near cosmic-ray sources would be detected as a single 
or cluster of point sources, depending on their angular scale.

\begin{figure}
\resizebox{7cm}{5cm}{\includegraphics{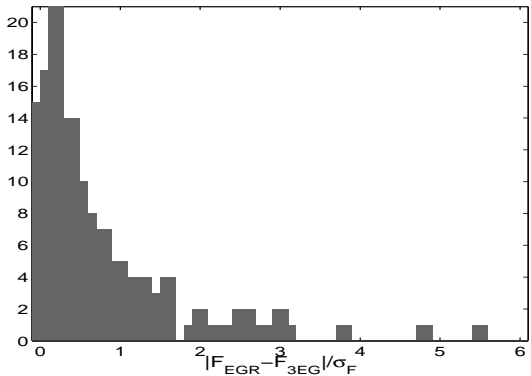}}
\centering \caption{Histogram of the relative flux differences $|F_{EGR}
- F_{3EG}| / \sigma_{F_{EGR}}$ measured between the EGR and 3EG counterparts in units of the statistical error on flux for each source. All fluxes are measured above 100 MeV.} \label{FigEGR3EGflux}
\end{figure}

\begin{figure}
\resizebox{7cm}{5cm}{\includegraphics{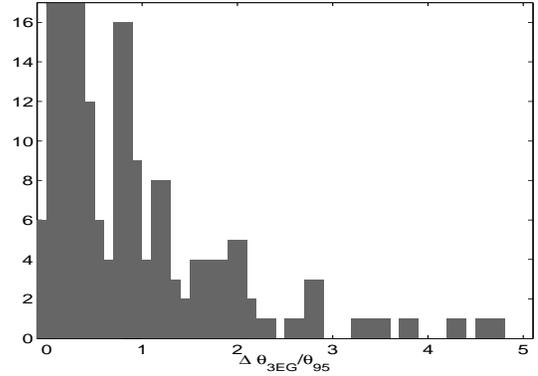}}
\centering \caption{Histogram of the relative angular separation between
the positions found for the EGR and 3EG counterparts in units of the 95\% confidence angle for each source.} 
\label{FigEGR3EGpos}
\end{figure}

For the 81 EGR sources that do have a 3EG counterpart, we find a
reasonable agreement in position and flux from both analyses. On
average, we find 3\% lower fluxes in the EGR analysis with respect
to the 3EG one because of the increase in Galactic background.
Figure \ref{FigEGR3EGflux} shows the histogram of ratios of the EGR and 3EG 
flux difference over the statistical error on flux for each source:  $|F_{EGR}- F_{3EG}| / \sigma_{F_{EGR}}$. 
The EGR flux was taken for the observation with highest $\sqrt{TS}$ and compared to the 3EG counterpart flux
for the same time period if available. Average P19 fluxes were
compared to the 3EG P1234 average for non flaring sources. The
flux differences are modest (17\% rms dispersion) and in most cases smaller
than the statistical uncertainties on flux estimates. Similarly, Figure \ref{FigEGR3EGpos} shows that the angular
separations between EGR and 3EG counterparts are often consistent with
the $\theta_{95}$ error radii. Yet,
thirty sources have been found as far as $0.5^{\circ}$ from the 3EG position and this will greatly
impact counterpart searches and identification at other
wavelengths.

On the other hand, we find 30 new EGR sources with no 3EG
counterpart. Their names are listed in Table \ref{TabEGRnew} and
they are displayed in Figure \ref{FigskyEGRnew}. Most of them are
detected just above the threshold and 11 of them were indeed
present in the 3EG complementary list, just below the significance
threshold.

\section{EGR source distributions and potential counterparts}

\begin{figure}
\resizebox{7cm}{5cm}{\includegraphics{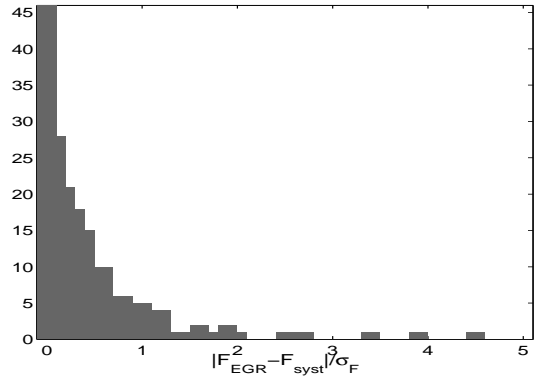}}
\centering \caption{Histogram of the relative flux differences 
$|F_{EGR}-F_{sys}|/ \sigma_{F_{EGR}}$ measured with the Ring and 
Galprop models in units of the statistical error on flux for each source. All fluxes are measured above 100 MeV.} \label{FigEGRsystflux}
\end{figure}

\begin{figure}
\resizebox{7cm}{5cm}{\includegraphics{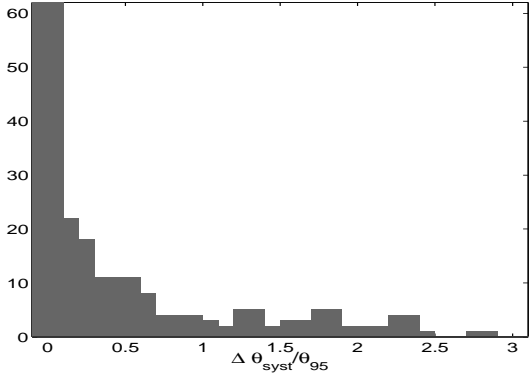}}
\centering \caption{Histogram of the relative angular separation between
the positions found with the Ring and Galprop models in units of the 95\% confidence angle for each source.} \label{FigEGRsystpos}
\end{figure}

Because of the new gas data we have used at intermediate latitude, the comparison between the EGR and 3EG source characteristics allows to judge, to some extent, the impact of our limited knowledge of gas mass tracers. The comparison between the flux and positions obtained with the Ring and Galprop models gives an estimate of the systematic
uncertainties due to our limited knowledge of the true cosmic-ray
distribution across the Galaxy. Figure \ref{FigEGRsystflux}  and Figure \ref{FigEGRsystpos} show that, in most cases, the differences are smaller than the statistical uncertainties. The distribution of 95\% confidence radii peaks between $\sim 0.2^{\circ}$  and $\sim 0.7^{\circ}$. The uncertainty in the background induces an additional
systematic error of $\sim 0.2^{\circ}$ for most sources. It should
be kept in mind while looking for counterparts.

\begin{figure*}
\includegraphics{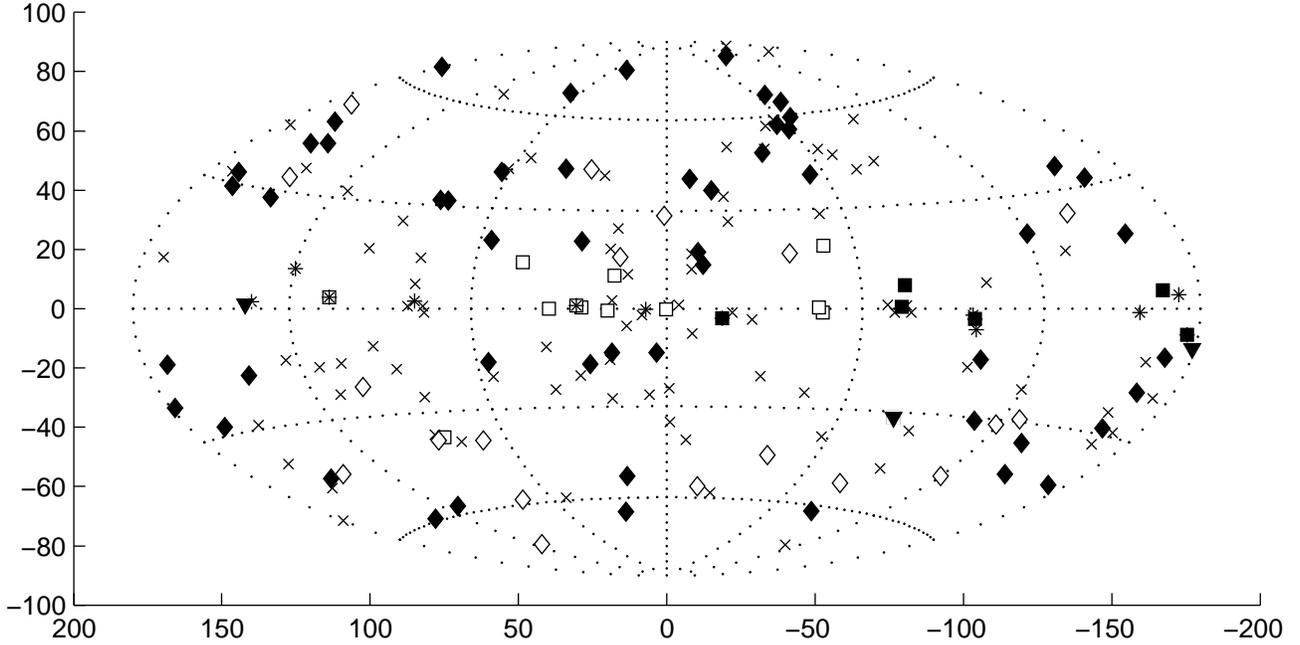}
\centering \caption{The revised EGRET source catalog, shown in
Galactic coordinates. The symbols indicate the counterpart types found in the error box: identified pulsars as black
squares; other ATNF pulsars as open squares; LSI +61 303, LMC,
and solar flare as black triangles; ASDC and CGRaBS blazar candidates as black diamonds; other
flat-spectrum radio-sources from CRATES as open diamonds; supernova remnants
from the Green catalogue as stars; no counterpart as crosses.} \label{FigskyEGRtypes}
\end{figure*}

\begin{figure}
\resizebox{7cm}{5cm}{\includegraphics{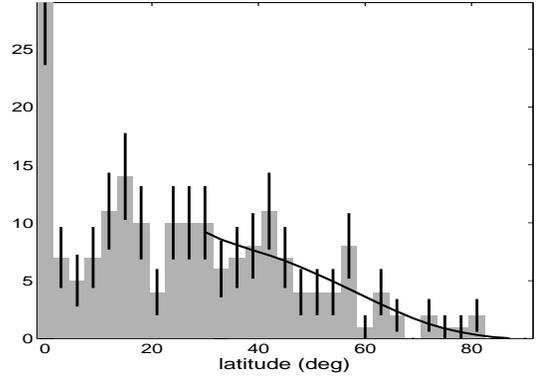}}
\centering \caption{Latitude distribution of the EGR sources with young Galactic sources at $|b| <3^{\circ}$, nearly
isotropically distributed sources far from the plane, as expected
from the black curve, and a flattening at mid-latitude because of
the rapid increase in the interstellar background flux.}
\label{FigEGRlatitude}
\end{figure}

We have searched the EGR error circles for potential counterparts
of interest such as pulsars from the ATNF catalogue (\cite{manchester05}), 
blazar candidates from the ASDC list (\cite{massaro05}) and the CGRaBS list (\cite{healey08}), other
flat radio sources from the CRATES compilation (\cite{healey07}), supernova remnants from the Green
catalogue (\cite{green06}), OB associations (\cite{Melnik95}), and X-ray and TeV pulsar wind nebulae (\cite{Li08} and \cite{grenier08}). The results are displayed in Figure
\ref{FigskyEGRtypes}. We have found 13 radio pulsar associations in addition to the 6 objects firmly identified by EGRET. 
Thirteen EGR sources coincide with supernova remnants, 9 with pulsar wind nebulae, 7 with OB associations, 53 with blazar candidates, and 19 with other flat radiosources. These associations should not be considered
as identification, but as spatial coincidences worthy of further investigations, in particular with the improved angular resolution of GLAST. Yet, they reveal that as many as 87 sources have no obvious counterpart among the
well-known $\gamma$-ray emitters despite the large number of
pulsars (1775) and radiosources (11 000) that were
cross-correlated with the sources and that spread across the entire sky and along the Galactic plane. The lack of blazar counterparts
is all the more surprising that the spatial distribution of the sources
off the plane is quite reminiscent of an isotropic, therefore extragalactic, distribution.
The latitude distribution, shown in Figure \ref{FigEGRlatitude},
is quite consistent above $30^{\circ}$ with a sample drawn from a uniform population according to the
exposure map, as shown by the black curve. The distribution
flattens at lower latitude because of the increased background
that drastically limits the survey sensitivity. Studying the
consistency with an extragalactic population at medium latitudes and the implication of the lack of a flat radiosource
is beyond the scope of this paper and will be addressed in a separate one. The sharp peak below $3^{\circ}$ in latitude
indicates young emitters. Their clustering in the inner Galaxy
($l \leq 30^{\circ}$), toward the direction tangent to the
Carina arm, and toward the Cygnus region outlines their close
relationship to large molecular complexes and star forming regions at a distance of a few kpc. 

\section{Discussion on specific sources}

\begin{figure}
\resizebox{8cm}{8cm}{\includegraphics{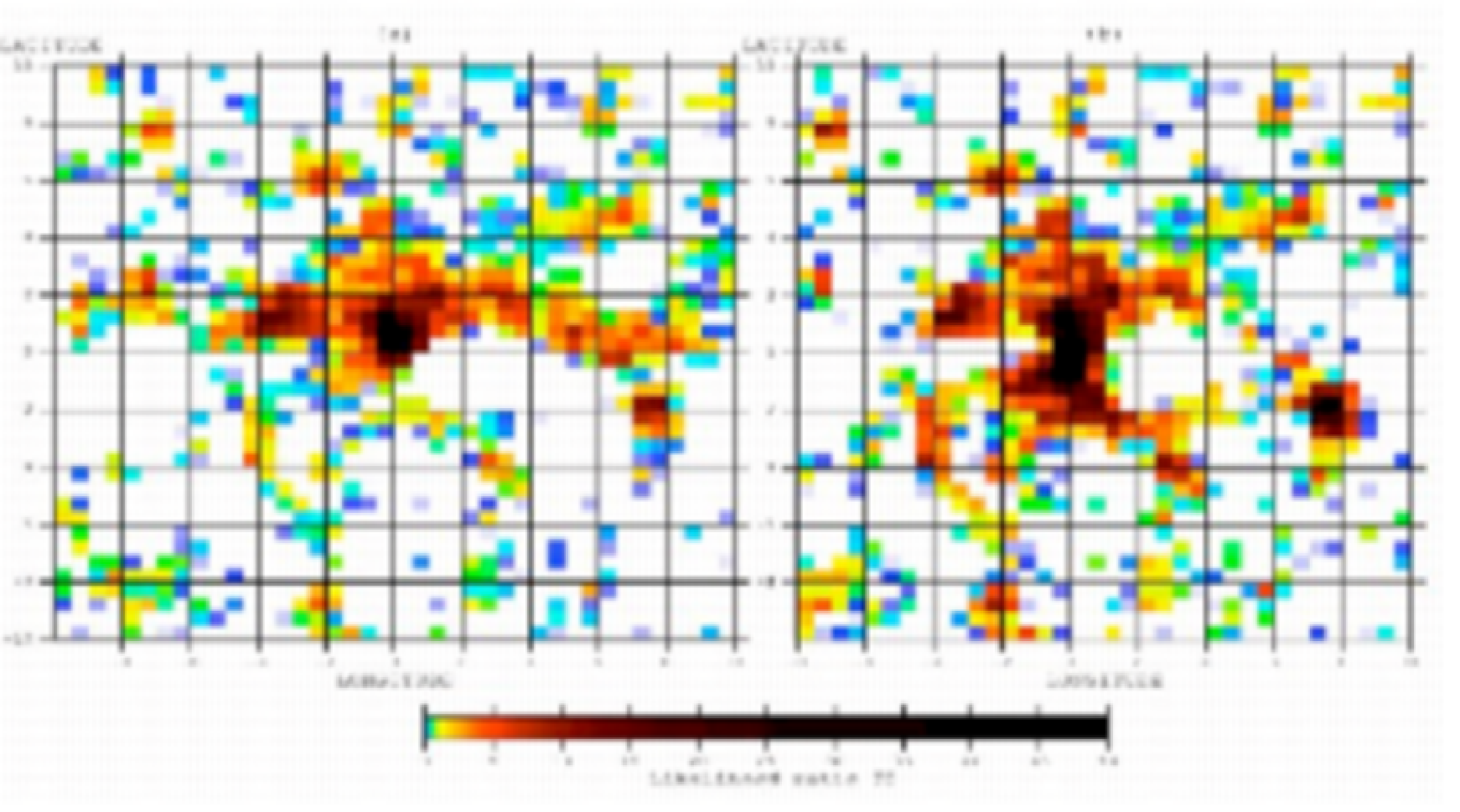}}
\centering \caption{$TS$-map obtained at energies above 1 GeV toward 
the Galactic center above the 3EG (a) and Ring (b) interstellar model} \label{FigGalCenter}
\end{figure}

There is considerable interest in the physical processes occurring in the Galactic center region. The 3EG catalogue lists one source located toward the Galactic center, 3EG J1744-3011. We find two point sources in this region, EGR J1740-2851 at $l = -0.55^{\circ}$, $b = 1.05^{\circ}$ and EGR J1747-2852 at $l = 0.21^{\circ}$, $b = -0.24^{\circ}$.
Figure \ref{FigGalCenter} display the $TS$-map for photons
with energies above 1 GeV above the 3EG and the Ring background models. The $\theta_{95}$ error radius around EGR J1740-2851 and EGR J1747-2852  formerly excludes the Galactic Center but source locations and fluxes in this direction should be taken with extreme caution
since the high gas optical depth around the Galactic center and the velocity pile-up toward the center induce large
uncertainties in the total gas column densities.

Coincidences with supernova remnants were noted (\cite{Sturner95}) and are confirmed in the present analysis (see Table \ref{TabEGRsnr}), but several also host a pulsar wind nebula, as in CTA 1 and IC 443, so we need much higher resolution $\gamma$-ray images to identify the origin of the emission, especially in these crowded regions. EGRET detections are confirmed toward two TeV-emitting wind nebulae around PSR J1420-6048 (in Kookaburra, EGRJ1418-6040) and PSR J1826-1334 (EGRJ1825-1325). Another interesting candidate is the wind nebula of the 11-kyr old and very energetic pulsar PSR J2229+6114 toward EGRJ2227+6114.

 \begin{table}
\caption{Names of the sources and supernova remnants found in spatial coincidence}
\label{TabEGRsnr} \centering
\begin{tabular}{l l l}     
\hline
EGRJ0008+7308     &G119.5+10.2  &CTA1\\
EGRJ0617+2238     &G189.1+3.0    &IC443\\
EGRJ0633+0646     &G205.5+0.5    &Monoceros\\
EGRJ1710-4435      &G343.0-6.0     &RCW114\\
EGRJ1800-2328      &G6.4-0.1          &W28\\
EGRJ1800-2328      &G6.5-0.4\\      
EGRJ1838-0420      &G27.8+0.6\\      
EGRJ1838-0420      &G28.8+1.5\\      
EGRJ2020+4019     &G78.2+2.1      &$\gamma$ Cygni\\
EGRJ2227+6114     &G106.3+2.7\\      
EGRJ0225+6240     &G132.7+1.3    &HB3\\
\hline
\end{tabular}
\end{table}

\begin{figure}
\resizebox{8cm}{6cm}{\includegraphics{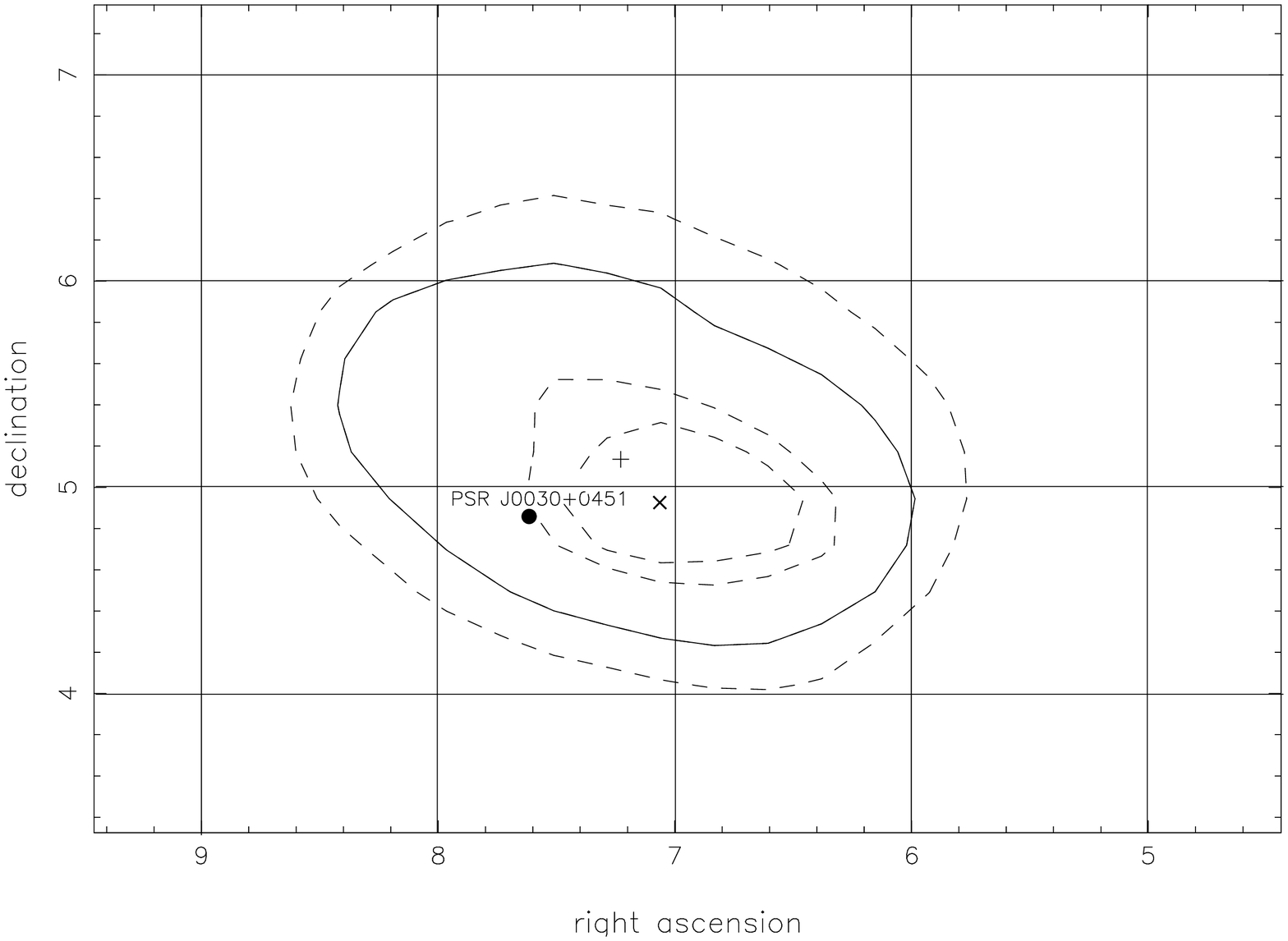}}
\centering \caption{Likelihood TS contours for energies 
above 100 MeV and periods incompassing PSR J0030+0451. The cross, the plus sign and the black dot respectively mark the EGR catalog position, the position with maximum likelihood and the pulsar location} \label{Fig0028}
\end{figure}

We also note, as shown in Figure \ref{Fig0028}, the positional
coincidence within 0.5$^{\circ}$ between the new EGR J0028+0457 source and the
millisecond X-ray pulsar PSR J0030+0451. This 300 pc distant pulsar, discovered in 2000 
 (\cite{somer2000}, \cite{damico2000}), has an X-ray counterpart exhibiting a double peaked 
pulse profile as seen by ROSAT (\cite{becker2000}). Millisecond pulsars 
have low magnetic fields, they 
produce relatively few electron-positron pairs so the electric field is not screened and 
the spectral cutoff due to pair production attenuation occurs at high energy.
They are therefore good candidate for accelerating particles to high energies. 
Harding et al. (2005) has predicted a $\gamma$-ray flux for PSR J0030+0451 well above the 
one of the $\gamma$-ray millisecond pulsar PSR J0218+4232 for which a pulsed emission 
was marginally detected (\cite{kuiper2000}). 

Four massive binaries have been detected at TeV energies, namely
PSR B1259-63 (\cite{aharonian05}), LSI +61$^{\circ}$ 303
(\cite{albert06}), LS 5039 (\cite{aharonian06}), and Cyg X-1
(\cite{albert07}), thus illustrating very efficient particle
acceleration in compressed or shocked pulsar winds, as well as in
microquasar jets. Inverse Compton scattering of the bright stellar radiation would dominate at GeV energies. We find no interesting EGRET counterpart to
these high-energy objects, but for the LSI +61$^{\circ}$ 303 radiosource.
The latter had long been associated with the COS-B source 2CG
135+01 and the EGRET source 2EG J0241+6119 (\cite{kniffen97}), yet
it had moved out of the 3EG error box and the marginal $\gamma$-ray
variability could not be associated with the radio flux
variations. In the present analysis, we find the radiosource very
near the centre of the EGR J0240+6112 source. On the other hand, we
find no source toward the dust enshrouded microquasar candidate,
AX J1639.0-4642, or the Be/X-ray binary, AO 0535+26, both proposed
as 3EG counterparts (\cite{combi03}, \cite{romero01}).

\begin{figure}
\resizebox{8cm}{6cm}{\includegraphics{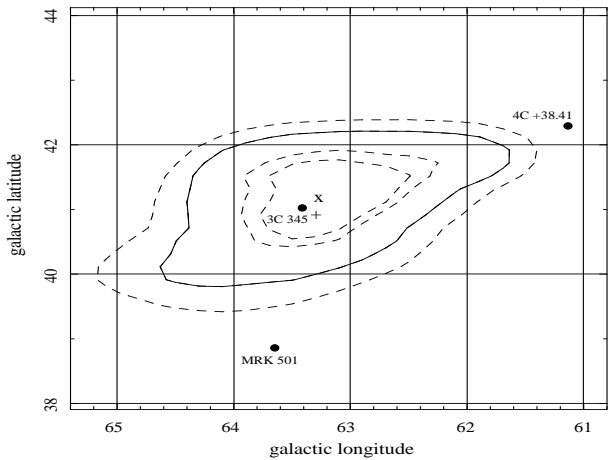}}
\centering \caption{Likelihood TS contours ( 50\%, 68\%, 
95\% and 99\% confidence) for energies above 100 MeV and period 5190. 
The cross is the EGR catalog position, the plus sign the position 
with maximum likelihood and the black dots mark the radio positions of 3C345, Mrk 501, and 4C+38.41} 
\label{Fig3C345}
\end{figure}

Another noticeable new source is EGR J1642+3940 detected 
at 5.8$\sigma$ rather close to 3C345.
3C345 is one of the most prominent flat spectrum ($\alpha=-0.1$) radio-loud, superluminal sources and is
therefore an excellent candidate for a $\gamma$-ray blazar. EGRET has viewed this region 12 times, in particular during period 5190 when a flare was found. We have analyzed again this particular period with the Ring model since it had not been used in the overall detection search. Figure \ref{Fig3C345} shows the
resulting TS contour for photons with energies above
100 MeV that is well centered on 3C345. The cross 
corresponds to the EGR position (period 5190), the plus sign to the 
position with maximum likelihood and the black dots to the position of 3C345 and a nearby AGN.  A marginal detection was also obtained for period 3034 at a level of 2.1$\sigma$. It should, however, be noted that the small photon excess above 500 MeV has been attributed to a flare from Mrk 501 by Kataoka et al. (1999) because the centroid was closer to the famous TeV source, so the association of EGR J1642+3940 with 3C345 is not clear. GLAST should easily confirm or infirm the association.

Several radiogalaxies (Cen A, NGC 6251, J1737-15) and a Seyfert 1
(GRS 1734-292) had been proposed as possible counterparts to
3EG sources (\cite{hartman99,combi03,foschini05,dicocco04}). They
triggered some interest because their identification would raise
important questions about the origin of the $\gamma$ rays at large angle from the strongly beamed emission from
the jet. We do not, however, confirm the spatial coincidence with EGR sources in
the present work. All these galaxies lie well beyond the 95\%
confidence region of EGR sources.

\section{Conclusions}
We have searched for point-like sources in the reprocessed EGRET data from cycle 1 to 9 
using new interstellar background models based on the most recent $HI$, $CO$, and dark gas data, 
as well as two different assumptions for the cosmic-ray distribution (the GALPROP diffusion 
model or a radial emissivity gradient fitted to the diffuse EGRET data). 
We have used the 3EG tools, likelihood method, procedure and significance threshold to detect 
sources, but have expanded the search to 3 different energy bands (above 100 MeV, 0.3-1 GeV, 
and above 1 GeV). The resulting number of detected sources has decreased by more than a third. 
Many unidentified sources, in particular among those spatially associated with the Gould Belt, 
are not confirmed as significant excesses. Their emission can be explained by the additional 
interstellar emission and its structure. Several interesting counterparts to 3EG sources, 
such as radiogalaxies, massive binaries, and microquasars, are now found outside the 95\% confidence region. 
We have cross-correlated the new source positions with large pulsar, supernova remant, pulsar wind nebulae, OB associations,
and radiosource catalogues, yet half the sample has no attractive counterpart among the 
potential $\gamma$-ray emitters. 30 new possible $\gamma$-ray sources have also been found.

This EGR catalog will be available in fits format at the Strasbourg astronomical Data Center (CDS) and in  
ASCII format at http://www.aim.univ-paris7.fr/EGRET\_catalogue/home.html 

\begin{acknowledgements}
We are deeply grateful to Bob Hartman for his helpful explanations about
the construction of the 3EG catalogue, and to Seth Digel and Andy
Strong for their help with the gas and Galprop maps.
\end{acknowledgements}


\Online

\begin{appendix} 
\end{appendix}

\begin{appendix} 
\tiny
\begin{longtable}{l l l l l l l l l l l l l l l l}
\caption{The EGR catalogue}\\ 
\label{TabEGR_full} 
\centering
Num&Name        & RA    & Dec   & l      & b     & $\theta_{95}$& F  & $\sigma_F$ &Cnts& $\sqrt{TS}$& vp & l$_{sys}$ & b$_{sys}$ & F$_{sys}$ & 3EG      \\
  1 &EGR J0008+7308&   2.01&  73.14&  119.75&  10.54& 0.20&  39.7&   4.4&  330&   10.9&     p19&  119.75&  10.54&   41.0&  3EGJ0010+7309\\
    &             &       &       &        &       &     &  63.9&  11.6&   96&    7.2&      p1&	   &	   &	   &		   \\
    &             &       &       &        &       &     &  33.4&   9.6&   61&    4.1&      p2&	   &	   &	   &		   \\
    &             &       &       &        &       &     &  22.4&   8.7&   37&    3.0&      p4&	   &	   &	   &		   \\
    &             &       &       &        &       &     &  48.8&   7.4&  162&    8.2&     p12&	   &	   &	   &		   \\
    &             &       &       &        &       &     &  21.6&   7.3&   52&    3.4&     p34&	   &	   &	   &		   \\
\hline
\end{longtable}
\end{appendix} 

\begin{appendix} 
\tiny
\begin{longtable}{l l l l l l l l l l l l l l l l l}
\caption{The EGR confused sources catalogue}\\ 
\label{TabEGRc} 
\centering
Num & Name        & RA    & Dec   & l      & b     & $\theta_{95}$& F  & $\sigma_F$ &Cnts& $\sqrt{TS}$& vp & l$_{sys}$ & b$_{sys}$ & F$_{sys}$ & 3EG & EGR     \\
1   &EGRc J0225+6240&  36.38&  62.68&  133.49&   1.75& 0.34&  30.2&   5.0&  344&    6.5&     p19&  133.05&   1.64&   22.1&  3EGJ0229+6151&  EGR J0240+6112\\
    &             &       &       &        &       &     &  42.3&  11.5&  100&    4.1&      p1&	   &	   &	   &		   &		  \\
    &             &       &       &        &       &     &  24.9&  10.6&   64&    2.5&      p2&	   &	   &	   &		   &		  \\
    &             &       &       &        &       &     &  23.9&  10.2&   63&    2.5&      p3&	   &	   &	   &		   &		  \\
    &             &       &       &        &       &     &  38.1&  14.4&   55&    2.9&      p4&	   &	   &	   &		   &		  \\
    &             &       &       &        &       &     &  35.1&   7.8&  173&    4.9&     p12&	   &	   &	   &		   &		  \\
\hline
\end{longtable}
\end{appendix} 

\end{document}